\documentclass[11pt]{article}
\usepackage{amsmath,amsfonts,amssymb,graphicx,epsfig,pdflscape,multirow,theorem,tikz,tikz-cd}
\usepackage{simplewick}
\usetikzlibrary{decorations.markings}
\tikzset{middlearrow/.style={
        decoration={markings,
            mark= at position 0.5 with {\arrow{#1}} ,
        },
        postaction={decorate}
    }
}
\numberwithin{equation}{section}
\graphicspath{{./eps/}}
\usepackage{braket}
\usepackage{mathtools}

\long\def\ignore#1{}
\definecolor{darkblue}{rgb}{0,0,.8}
\definecolor{red}{rgb}{1,0,0}
\definecolor{purple}{rgb}{1,0,1}
\definecolor{coloroflink}{rgb}{0.7,0,1}
\definecolor{coloroflink}{rgb}{0.180392, 0.545098, 0.341176}
\definecolor{darkpurple}{rgb}{1,.2,1}
\definecolor{pink}{rgb}{1,.7,.7}
\definecolor{lightblue}{rgb}{.61,.61,1}
\definecolor{midblue}{rgb}{.7,.7,1}
\definecolor{lightlightblue}{rgb}{.9,.9,1}
\definecolor{lightestblue}{rgb}{.96,.96,1}
\definecolor{lightpurple}{rgb}{1,.65,1}
\definecolor{darkgreen}{rgb}{0.180392, 0.545098, 0.341176}

\theorembodyfont{\itshape} 
\theoremheaderfont{\scshape}
\theoremstyle{plain}

\numberwithin{equation}{section}

\newcommand{\nc}{\newcommand}
\nc{\bib}{\bibitem}
\nc{\be}{\begin{equation}}
\nc{\ee}{\end{equation}}
\nc{\nn}{\nonumber\\ }
\nc{\chit}{\raisebox{0.25ex}{$\chi$}}
\nc{\chih}{\raisebox{0.25ex}{$\hat\chi$}}

\nc{\g}{\mathfrak{g}}
\nc{\gh}{\widehat{\mathfrak{g}}}
\nc{\Ac}{\mathcal{A}}
\nc{\Bc}{\mathcal{B}}
\nc{\Acb}{\bar{\mathcal{A}}}
\nc{\Ic}{\mathcal{I}}
\nc{\Mc}{\mathcal{M}}
\nc{\Nc}{\mathcal{N}}
\nc{\Oc}{\mathcal{O}}
\nc{\Qc}{\mathcal{Q}}
\nc{\Vc}{\mathcal{V}}
\nc{\Vir}{\mathfrak{Vir}}
\nc{\Virb}{\overline{\mathfrak{Vir}}}
\nc{\pa}{\partial}
\nc{\eps}{\epsilon}
\nc{\Tb}{\bar{T}}
\nc{\Wb}{\bar{W}}
\nc{\cb}{\bar{c}}
\nc{\Ab}{\bar{A}}
\nc{\Bb}{\bar{B}}
\nc{\Cb}{\bar{C}}
\nc{\Jb}{\bar{J}}
\nc{\zb}{\bar{z}}
\nc{\kb}{\bar{k}}
\nc{\C}{\mathbb{C}}
\nc{\al}{\alpha}
\nc{\bet}{\beta}
\nc{\WA}{W\!A}

\nc{\Ah}{\widehat{A}}
\nc{\Bh}{\widehat{B}}
\nc{\Ch}{\widehat{C}}
\nc{\Jh}{\widehat{J}}
\nc{\Th}{\widehat{T}}
\nc{\Uh}{\widehat{U}}
\nc{\Vh}{\widehat{V}}
\nc{\Wh}{\widehat{W}}

\nc{\ii}{\mathrm{i}}

\nc{\gast}{\!\ast}
\nc{\s}{\;\!\!}

\nc{\La}{\Lambda}
\nc{\Lh}{\hat{\Lambda}}

\nc{\I}{\mathbb{I}}

\newcommand\bea{\begin{eqnarray}}
\newcommand\eea{\end{eqnarray}}

\begin{document}

\topmargin -15mm
\oddsidemargin 05mm

%
%

\title{\mbox{}\vspace{0in}
\bf 
\huge
Higher-order Galilean contractions
\\[-.3cm]
}
\date{}

\maketitle


\begin{center}
{\vspace{-5mm}\LARGE J{\o}rgen Rasmussen,\, Christopher Raymond}
\\[.5cm]
{\em School of Mathematics and Physics, University of Queensland}\\
{\em St Lucia, Brisbane, Queensland 4072, Australia}
\\[.4cm] 
{\tt j.rasmussen\,@\,uq.edu.au}
\qquad
{\tt christopher.raymond\,@\,uqconnect.edu.au}
\end{center}

%
%

\vspace{0.5cm}
\begin{abstract}
A Galilean contraction is a way to construct Galilean conformal algebras from a pair of infinite-dimensional conformal algebras, or equivalently, a method for contracting tensor products of vertex algebras. Here, we present a generalisation of the Galilean contraction prescription to allow for inputs of any finite number of conformal algebras, resulting in new classes of higher-order Galilean conformal algebras. We provide several detailed examples, including infinite hierarchies of higher-order Galilean Virasoro algebras, affine Kac-Moody algebras and the associated Sugawara constructions, and $W_{3}$ algebras.
\end{abstract}


\newpage
\section{Introduction}
\label{Sec:Introduction}

The Galilean Virasoro algebra appears in studies of asymptotically flat three-dimensional spacetimes, 
see \cite{BO14} and references therein.
It can be constructed \cite{BC07,BG09,HR10,BGMM10,BF12} as an In\"on\"u-Wigner contraction \cite{Seg51,IW53,Sal61,Gil06} 
of a commuting pair of Virasoro algebras. The Galilean $W_3$ algebra \cite{ABFGR13,GMPT13,ChrisThesis,RR17} likewise follows
by contracting a pair of $W_3$ algebras \cite{Zam85}.
Many other Galilean conformal algebras with extended symmetries have been worked out \cite{GRR15,ChrisThesis,RR17}, 
including contractions of higher-rank $W_N$ algebras \cite{FL88,BFKNRV91,KW91,Hor93,Zhu94}.
Earlier works contributing to our understanding of non-relativistic systems with (typically non-affine) conformal 
symmetry can be found in \cite{Hag72,Nie72,Hen94,NORM97,LSZ06,NS07,DH09}.

The constructions of the (affine) conformal algebras are all based on contractions of {\em pairs} of symmetry algebras, 
or equivalently, contractions of tensor products of two vertex algebras.
In this note, we present a generalisation to allow for inputs of any finite number of symmetry algebras.
In the general construction, these algebras are all assumed identical, up to their central charges, although asymmetric
contractions are possible, as we discuss briefly.
These results solidify ideas put forward in \cite{RR17} and give rise to new infinite hierarchies of higher-order Galilean conformal algebras.

Higher-order Galilean Virasoro algebras have since appeared in work \cite{CCRSR17} on
the so-called $S$-expansion method \cite{IRS06}.
Moreover, in a recent study \cite{CMRSRV19} of a non-abelian enlargement of the Poincar\'e symmetry algebra associated with
a Chern-Simons theory on AdS$_3$, the flat-space asymptotics of the ensuing AdS-Lorentz symmetry algebra \cite{SS09}
has been found to give rise to a third-order Galilean Virasoro algebra.
As it has also been found \cite{BH86,HenRey10,CFPT10,CFP11,GG11} that the asymptotic algebra of higher-spin gravity on AdS$_3$ 
exhibits a $W$-symmetry, it thus seems natural to expect that higher-order Galilean $W$-algebras will play a role in higher-spin 
Chern-Simons models on AdS spacetimes with enlarged Poincar\'e symmetry.

In Section \ref{Sec:Galilean}, we outline the generalised contraction prescription and illustrate it by 
working out the higher-order Galilean Virasoro and affine Kac-Moody algebras.
In Section \ref{Sec:GenSug}, we construct a Sugawara operator \cite{Sug68} for each Galilean Kac-Moody algebra; 
its central charge is given by the product of the contraction order and the dimension of the underlying Lie algebra.
We also show that the Sugawara construction commutes with the Galilean contraction procedure.
In Section \ref{Sec:GW3}, we apply the Galilean contractions to the $W_3$ algebra and thereby obtain an infinite hierarchy 
of higher-order $W_3$ algebras.
Section \ref{Sec:Discussion} contains some concluding remarks.

\section{Galilean contractions}
\label{Sec:Galilean}

\subsection{Operator-product algebras and star relations}
\label{Sec:OPA}

It is often convenient to combine the generators of the
symmetry algebra of a conformal field theory into generating fields of the form
\be
A(z) = \sum_{n\in-\Delta_A + \mathbb{Z}} A_n\,z^{-n-\Delta_{A}},
\ee
where $\Delta_{A}$ is the conformal weight of $A$. We are interested in the corresponding operator-product algebra (OPA) $\mathcal{A}$,
where the operator-product expansion (OPE) of the two fields $A,B\in\Ac$ is given by
\be
A(z)B(w) = \sum^{\Delta_{A} + \Delta_{B}}_{n=-\infty} \frac{[AB]_{n}(w)}{(z-w)^{n}}.
\ee
Here, if nonzero, $[AB]_{n}$ is a field of conformal weight $\Delta_{A} + \Delta_{B} - n$.
As the nontrivial information of an OPE is stored in the singular terms, one often ignores the non-singular terms, writing
\be
 A(z)B(w) \sim \sum^{\Delta_{A} + \Delta_{B}}_{n=1} \frac{[AB]_{n}(w)}{(z-w)^{n}}.
\label{OPEsing}
\ee
The normal ordering of $A,B\in\Ac$ is given by $(AB)=[AB]_0$. We use $\I$ to denote the identity field.

An OPA $\Ac$ is said to be conformal if it contains a distinct field $T$ generating a Virasoro subalgebra. 
In that case, a field $A \in \Ac$ is called a scaling field if
\be
[TA]_{2} = \Delta_{A} A, \qquad [TA]_{1} = \partial A.
\ee
Such a field is {\em quasi-primary} if $[TA]_{3} = 0$, and {\em primary} if $[TA]_{n}= 0$ for all $n \geq 3$. 
Let $\Bc_{\Ac}$ denote a basis for the linear span of the quasi-primary fields in $\Ac$. 
Relative to this, the OPE (\ref{OPEsing}) reads
\be
 A(z)B(w) \sim \sum_{Q\in\Bc_{\Ac}} C_{A,B}^{Q} \left( \sum_{n=0}^{\Delta_{A} + \Delta_{B} - \Delta_{Q}} 
  \frac{\beta^{\Delta_{Q};n}_{\Delta_{A},\Delta_{B}}\partial^{n}Q(w)}{(z-w)^{\Delta_{A} + \Delta_{B} - \Delta_{Q} - n}} \right),
\label{AB}
\ee
with structure constants $C^{Q}_{A,B}$ and 
\be
\beta^{\Delta_{Q};n}_{\Delta_{A},\Delta_{B}} = \frac{(\Delta_{A} - \Delta_{B} + \Delta_{Q})_{n}}{n!(2\Delta_{Q})_{n}},
\qquad (x)_{n} = \prod_{j=0}^{n-1} (x+j).
\ee
Compactly, we may represent the OPE (\ref{AB}) by the so-called star relation
\be
 A \ast B \simeq \sum_{Q\in\Bc_{\Ac}} C_{A,B}^{Q} \{ Q\},
\ee
where $\{Q\}$ represents the sum over $n$.
We refer to \cite{RR17,Thi95} for more details on the algebraic structure of an OPA.

\subsection{Contraction prescription}
\label{Sec:Higher}

For $N\in\mathbb{N}$, we consider the tensor-product algebra
\be
 \Ac^{\otimes N}=\bigotimes_{i=0}^{N-1}\Ac_{(i)},
\ee
where, for simplicity, $\Ac_{(0)},\ldots,\Ac_{(N-1)}$ are copies of the same OPA $\Ac$, up to the value of their central parameters
(such as central charges).
For $\eps\in\mathbb{C}$, let
\be
 A_{i,\eps}=\eps^i\sum_{j=0}^{N-1}\omega^{ij}A_{(j)}, \qquad
 c_{i,\eps}=\eps^i\sum_{j=0}^{N-1}\omega^{ij}c_{(j)},
 \qquad i=0,\ldots,N-1,
\label{Aie}
\ee
where $A_{(j)}$ (respectively $c_{(j)}$) denotes the field $A\in\Ac_{(j)}$ (respectively the central parameter $c$), 
and $\omega$ is the principal $N$th root of unity, 
\begin{equation}
\omega=e^{2\pi\ii/N}.
\end{equation}
For $\eps \neq0$, the map
\be
 \Ac^{\otimes N}\to\Ac^{\otimes N},\qquad (A_{(0)},\ldots,A_{(N-1)})\mapsto(A_{0,\eps},\ldots,A_{N-1,\eps}),
\label{AA}
\ee
(and similarly for the central parameters) is invertible, with
\be
 A_{(i)}=\frac{1}{N}\sum_{j=0}^{N-1}\omega^{-ij}\eps^{-j}A_{j,\eps},\qquad i=0,\ldots,N-1.
\label{invbasis}
\ee

In the special case $N=2$, we have $\omega=-1$ and
\be
 A_{0,\eps}=A_{(0)}+A_{(1)},\qquad A_{1,\eps}=\eps\big(A_{(0)}-A_{(1)}\big),
\ee
with inverses
\be
 A_{(0)}=\tfrac{1}{2}\big(A_{0,\eps}+\tfrac{1}{\eps}A_{1,\eps}\big),\qquad
 A_{(1)}=\tfrac{1}{2}\big(A_{0,\eps}-\tfrac{1}{\eps}A_{1,\eps}\big).
\ee
In \cite{RR17}, these fields are denoted by
\be
 A=A_{(0)},\qquad \bar{A}=A_{(1)},\qquad A_\eps^+=A_{0,\eps},\qquad A_\eps^-=A_{1,\eps}.
\ee

For $\eps=0$, the map (\ref{AA}) is singular (unless $N=1$), indicating that a new algebraic structure emerges
in the limit $\eps\to0$, where 
\be
 A_{i,\eps}\to A_i,\qquad c_{i,\eps}\to c_i.
\ee
If the resulting algebra is a well-defined OPA, we refer to it as the {\em $N$th-order Galilean OPA} $\Ac^N_G$.
In particular, if $\Ac$ is an OPA of Lie type (that is, the underlying algebra of modes is a Lie algebra), 
then all the corresponding higher-order Galilean contractions are indeed well-defined and readily obtained.
This is illustrated by the Virasoro and affine Kac-Moody algebras in Section \ref{Sec:Virasoro}.

\subsection{Galilean Virasoro and affine Kac-Moody algebras}
\label{Sec:Virasoro}

The Virasoro OPA $\Vir$ of central charge $c$ is of Lie type and generated by $T$, with star relation
\be
T \ast T \simeq \tfrac{c}{2}\{\I\} + 2 \{ T \}.
\ee
The Galilean Virasoro algebra of order $N$, $\Vir^N_G$, is generated by the fields 
$T_0,\ldots,T_{N-1}$, with central parameters $c_0,\ldots,c_{N-1}$ and star relations
\be
T_i\ast T_j\simeq
\begin{dcases}
\tfrac{c_{i+j}}{2}\{\I\}+2\{T_{i+j}\}, \ &i+j < N, \\[.15cm] 
 0, \ &i+j \geq N.
 \end{dcases}
 \label{HOGalVir}
\ee
This yields an infinite family of extended Virasoro algebras, $\{\Vir_G^N\,|\,N\in\mathbb{N}\}$, where
$\Vir_G^1\cong\Vir$ while $\Vir_G^2$ is the familiar Galilean Virasoro algebra \cite{BC07,BG09,HR10,BGMM10,BF12,ChrisThesis,RR17}.
For small $N$, the Galilean Virasoro algebras $\Vir_G^N$ have recently appeared in \cite{CCRSR17}.

The OPE of two fields in an affine Kac-Moody (or current) algebra $\gh$ (where the central element $K$ has been
replaced by $k\,\I$, with $k$ the level) is given by
\be
 J^a(z)J^b(w)\sim\frac{\kappa^{ab}k}{(z-w)^2}+\frac{{f^{ab}}_cJ^c(w)}{z-w},
\ee
where ${f^{ab}}_c$ are structure constants and $\kappa$ the Killing form of the underlying finite-dimensional Lie algebra $\g$.
(As is customary, the summation over the basis label $c$ is not displayed.)
The corresponding OPA is of Lie type, 
and we find that $\gh_G^{\,N}$ is generated by $\{J_i^a\,|\,a=1,\ldots,\dim\g;\,i=0,\ldots,N-1\}$, with nontrivial star relations
\be
 J^a_i\ast J^b_j\simeq\kappa^{ab}k_{i+j}\{\I\}+{f^{ab}}_c\{J^c_{i+j}\},\qquad i+j\in\{0,\ldots,N-1\}.
\ee

In the limit $N\to\infty$, we obtain the algebra $\gh_G^{\,\infty}$; it is generated by $\{J_i^a\,|\,a=1,\ldots,\dim\g;\,i\in\mathbb{N}_0\}$,
with nontrivial star relations
\be
 J^a_i\ast J^b_j\simeq\kappa^{ab}k_{i+j}\{\I\}+{f^{ab}}_c\{J^c_{i+j}\}.
\ee
It follows that
\be
 \gh_G^{\,\infty}\,\cong\,\gh\otimes\mathbb{C}[t]
\ee
and
\be
 \gh_G^{\,N}\cong\gh\otimes\mathbb{C}[t]/\langle t^N\rangle,
\ee
extending to general $N$ the construction of the Takiff algebras considered in \cite{HSSU12,BR13}.
We similarly have
\be
 \Vir_G^{\infty}\,\cong\,\Vir\otimes\mathbb{C}[t],\qquad \Vir_G^N\cong\Vir\otimes\mathbb{C}[t]/\langle t^N\rangle.
\ee

\section{Generalised Sugawara constructions}
\label{Sec:GenSug}

In~\cite{RR17}, we constructed a Sugawara operator for Galilean affine Kac-Moody algebras (of order $2$), 
and showed that this process commutes with the Galilean contraction procedure.
We find that a similar result holds for the higher-order Galilean affine Kac-Moody algebras, manifested by the commutativity of the diagram
\begin{center}
\begin{tikzcd}[column sep=large,row sep = large]
 \gh^{\,\otimes N} \arrow[r, "\mathrm{Sug}^{\otimes N}"] \arrow[d,two heads, "\mathrm{Gal}"]
&\Vir^{\otimes N} \arrow[d,two heads, "\mathrm{Gal}"] \\
\gh^{\,N}_G \arrow[r, "\mathrm{Gal\; Sug}"]
& \Vir^N_G
\end{tikzcd}
\end{center}
To verify this, separate analyses of the two branches are presented in the following two subsections:
The lower branch is considered in Section \ref{Sec:ConSug}; the upper one in Section \ref{Sec:SugCon}.

\subsection{Galilean Sugawara construction}
\label{Sec:ConSug}

For the generators of $\Vir_G^N$, we make the ansatz
\be
 T_i=\sum_{r,s= 0}^{N-1}\lambda^{r,s}_i \kappa_{ab}(J_r^aJ_s^b),\qquad i=0,\ldots,N-1,
\label{Ti}
\ee
where $\kappa_{ab}$ are elements of the inverse Killing form on $\g$. The task is now to determine
the coefficients $\lambda^{r,s}_i$ such that
\be
 T_i\ast J_j^a\simeq\begin{dcases} \{J_{i+j}^a\},\ &i+j\in\{0,\ldots,N-1\},\\[.15cm] 0,\ &i+j\geq N.\end{dcases}
\label{TJ}
\ee
We show below that this is indeed possible. It subsequently follows that $\Vir_G^N=\langle T_0,\ldots,T_{N-1}\rangle$, with central charges
\be
 c_0=N\dim\g,\qquad c_1,\ldots,c_{N-1}=0.
\label{cSug}
\ee

First, we compute the OPE
\begin{align}
 J_j^a(z)T_i(w)&=\sum_{r,s=0}^{N-1}\frac{\lambda_i^{r,s}}{(z-w)^2}\big\{k_{j+r}J_s^a(w)+k_{j+s}J_r^a(w)+2h^\vee J_{j+r+s}^a(w)\big\}
 \nonumber\\[.15cm]
 &+\sum_{r,s=0}^{N-1}\frac{\lambda_i^{r,s}\kappa_{bc}}{z-w}\big\{{f^{ab}}_d(J_{j+r}^dJ_s^c)(w)
  +{f^{ac}}_d(J_r^bJ_{j+s}^d)(w)\big\},
\label{JT}
\end{align}
where $h^\vee$ is the dual Coxeter number of $\mathfrak{g}$, arising through the relation 
$\kappa_{bc}{f^{ab}}_d{f^{dc}}_e=2h^\vee\delta_e^a$.
To satisfy (\ref{TJ}), the first sum must equal $J_{i+j}^a(w)/(z-w)^2$ while the second sum must vanish.
The second-sum constraint implies that
\be
 \lambda_i^{r,s}=\begin{dcases}\lambda_i^{\ell,N-1},\ & r+s=N-1+\ell\quad (\ell=0,\ldots,N-1),\\[.15cm]
  0,\ & r+s<N-1.
  \end{dcases}
\ee
This leaves $N$ coefficients, $\lambda_i^{0,N-1},\ldots,\lambda_i^{N-1,N-1}$, for each $i\in\{0,\ldots,N-1\}$.
The first-sum constraint then requires that
\be
 2\sum_{n=j}^{N-1}\sum_{\ell=0}^{n-j}\lambda_i^{\ell,N-1}k_{N-1-n+j+\ell}J_n^a+2Nh^\vee\lambda_i^{0,N-1}\delta_{j,0}J_{N-1}^a
  =\begin{dcases} J_{i+j}^a,\ &i+j\leq N-1,\\[.15cm] 0,\ &i+j\geq N.\end{dcases}
\ee
For each $i$, this translates into a lower-triangular system of linear equations:
\be
 2\begin{pmatrix} 
 k_{N-1}&&&&\\ 
 k_{N-2}&k_{N-1}&&&\\[.15cm]
 \vdots&\ddots&\ddots&&\\
 k_1&&\ddots&k_{N-1}&\\
 k_0'&k_1&\cdots&k_{N-2}&k_{N-1}
 \end{pmatrix}
 \begin{pmatrix} \lambda_i^{0,N-1}\\ \vdots\\ \\ \vdots\\ \lambda_i^{N-1,N-1}\end{pmatrix}
 =\begin{pmatrix} 0\\ \vdots\\ 1\\ \vdots\\ 0\end{pmatrix},
\ee
where $k_0'=k_0+Nh^\vee$, and where the only nonzero component on the right-hand side is a $1$ in position $i+1$.
To solve these systems, we must assume that $k_{N-1}\neq0$, in which case the problem reduces to
inverting the lower-triangular Toeplitz matrix
\be
 A=\begin{pmatrix} 
 1&&&&\\[.15cm]
 a_1&1&&&\\
 a_2&\ddots&\ddots&&\\
 \vdots&\ddots&\ddots&\,1&\\[.1cm]
 a_{N-1}&\cdots&a_2&\,a_1&1
 \end{pmatrix},
\label{A}
\ee
where
\be
 a_m=\frac{k_{N-1-m}+Nh^\vee\delta_{m,N-1}}{k_{N-1}},\qquad m=1,\ldots,N-1.
\label{am}
\ee
The inverse is itself a lower-triangular Toeplitz matrix with $1$'s on the diagonal,
\be
 A^{-1}=\begin{pmatrix} 
 b_0&&&&\\[.15cm]
 b_1&b_0&&&\\
 b_2&\ddots&\ddots&&\\
 \vdots&\ddots&\ddots&\,b_0&\\[.1cm]
 b_{N-1}&\cdots&b_2&\,b_1&b_0
 \end{pmatrix},\qquad b_0=1,
\label{Ainv}
\ee
and we find that the nontrivial matrix elements are given by
\be
 b_n=\sum_{p\in(\mathbb{N}_0)^n}(-1)^{|p|}\frac{\delta_{|\!|p|\!|,n}|p|!}{p_1!\cdots p_n!}\,a_1^{p_1}\cdots a_n^{p_n},
\label{bn}
\ee
where
\be
 |p|=\sum_{i=1}^n p_i,\qquad |\!|p|\!|=\sum_{i=1}^ni p_i,\qquad p=(p_1,\ldots,p_n).
\ee
It follows that 
\be
 \lambda_i^{\ell,N-1}=\begin{dcases} 0,\ &\ell=0,\ldots,i-1,\\[.15cm]
   \frac{b_{\ell-i}}{2k_{N-1}},\ &\ell=i,\ldots,N-1,\end{dcases}
\ee
so the {\em unique} expression for $T_i$ of the form (\ref{Ti}) is given by
\be
 T_i=\sum_{n=0}^{N-1-i}\frac{b_n}{2k_{N-1}}\sum_{t=0}^{N-1-i-n}\kappa_{ab}(J_{i+n+t}^aJ_{N-1-t}^b).
\label{Tifinal}
\ee
For $N=2$, we thus recover the Galilean Sugawara construction obtained in \cite{RR17}, 
\be
 T_0=\frac{\kappa_{ab}}{2k_1}\{(J_0^aJ_1^b)+(J_1^aJ_0^b)\}
  -\frac{k_0+2h^\vee}{2(k_1)^2}\kappa_{ab}(J_1^aJ_1^b),\qquad 
 T_1=\frac{\kappa_{ab}}{2k_1}(J_1^aJ_1^b),
\ee
whereas for $N=3$, we find the new expressions
\begin{align}
 T_0&=\frac{\kappa_{ab}}{2k_2}\{(J_0^aJ_2^b)+(J_1^aJ_1^b)+(J_2^aJ_0^b)\}
  -\frac{k_1\kappa_{ab}}{2(k_2)^2}\{(J_1^aJ_2^b)+(J_2^aJ_1^b)\}
  +\frac{(k_1)^2-(k_0+3h^\vee)k_2}{2(k_2)^3}\kappa_{ab}(J_2^aJ_2^b),
 \nonumber\\[.15cm]
 T_1&=\frac{\kappa_{ab}}{2k_2}\{(J_1^aJ_2^b)+(J_2^aJ_1^b)\}
  -\frac{k_1\kappa_{ab}}{2(k_2)^2}(J_2^aJ_2^b),
 \qquad
 T_2=\frac{\kappa_{ab}}{2k_2}(J_2^aJ_2^b).
 \nonumber
\end{align}

For each $i=0,\ldots, N-1$, the value of the central parameter $c_{i}$ follows from the leading pole in the OPE $T_{0}(z)T_{i}(w)$. 
Using \eqref{Tifinal}, we compute
\be
  T_0(z)T_{i}(w) \sim\sum_{n=0}^{N-1-i}\frac{b_{n}}{2k_{N-1}}\sum_{t=0}^{N-1-i-n}\frac{\kappa_{ab}\kappa^{ab}k_{N-1+i+n}}{(z-w)^4}
  + \cdots,
\ee
suppressing all subleading poles.
Since $k_{a} = 0$ for $a \geq N$, this term is zero unless $n + i = 0$, that is, unless $n=i=0$. 
From $\kappa_{ab}\kappa^{ab} = \dim \mathfrak{g}$, we then obtain the announced result (\ref{cSug}).

\subsection{Sugawara before Galilean contraction}
\label{Sec:SugCon}

On the individual factors of $\gh^{\,\otimes N}$\!, the Sugawara construction is given by
\be
 T_{(i)}=\frac{\kappa_{ab}}{2(k_{(i)}+h^\vee)}(J_{(i)}^aJ_{(i)}^b),\qquad
 c_{(i)}=\frac{k_{(i)}\dim\g}{k_{(i)}+h^\vee},\qquad i=0,\ldots,N-1.
\ee
Changing basis as in (\ref{Aie}) introduces
\begin{align}
 T_{i,\eps}&=\eps^i\sum_{j=0}^{N-1}\omega^{ij}T_{(j)}
 =\eps^i\sum_{j=0}^{N-1}\omega^{ij}
  \frac{\sum_{\ell,\ell'=0}^{N-1}\omega^{-j(\ell+\ell')}\eps^{-\ell-\ell'}\kappa_{ab}(J_{\ell,\eps}^aJ_{\ell',\eps}^b)}{2N\big(
  \sum_{m=0}^{N-1}\omega^{-jm}\eps^{-m}k_{m,\eps}+Nh^\vee\big)}
 \nonumber\\[.15cm]
 &=\frac{1}{2Nk_{N-1,\eps}}\sum_{j,\ell,\ell'=0}^{N-1}(\omega^j\eps)^{N-1+i-\ell-\ell'}\frac{\kappa_{ab}(J_{\ell,\eps}^aJ_{\ell',\eps}^b)}{
  1+\sum_{m=1}^{N-1}a_{m,\eps}(\omega^j\eps)^m},
\label{Tie}
\end{align}
where
\be
 a_{m,\eps}=\frac{k_{N-1-m,\eps}+Nh^\vee\delta_{m,N-1}}{k_{N-1,\eps}},\qquad m=1,\ldots,N-1.
\ee
Now, using that a lower-triangular $N\times N$ Toeplitz matrix of the form (\ref{A}) decomposes as
\be
 A = I + a_{1} \eta + \cdots + a_{N-1}\eta^{N-1},
\label{Aeta}
\ee
where $I$ is the identity matrix and $\eta$ the $N\times N$ matrix
\be
 \eta=\begin{pmatrix} 
 0&&&&\\
 1&0&&&\\
 0&\ddots&\ddots&&\\[-.1cm]
 \vdots&\ddots&\ddots&0&\\
 0&\cdots&0&1&0
 \end{pmatrix},
\ee
we can use the result for $A^{-1}$ in (\ref{Ainv})-(\ref{bn}) to expand the expression for $T_{i,\eps}$ in powers of $\eps$.
We thus find that
\begin{align}
 T_{i,\eps}&=\frac{1}{2Nk_{N-1,\eps}}\sum_{j,\ell,\ell'=0}^{N-1}(\omega^j\eps)^{N-1+i-\ell-\ell'}\kappa_{ab}(J_{\ell,\eps}^aJ_{\ell',\eps}^b)
  \Big(\sum_{n=0}^{N-1}b_{n,\eps}(\omega^j\eps)^b+\mathcal{O}(\eps^N)\Big)
 \nonumber\\[.15cm]
 &=\frac{1}{2Nk_{N-1,\eps}}\sum_{\ell,\ell',n=0}^{N-1}b_{n,\eps}\kappa_{ab}(J_{\ell,\eps}^aJ_{\ell',\eps}^b)
  \sum_{j=0}^{N-1}(\omega^j\eps)^{N-1+i-\ell-\ell'+n}
  +\mathcal{O}(\eps^{i+1}),
\end{align}
where
\be
 b_{0,\eps}=1,\qquad b_{n,\eps}=\sum_{p\in(\mathbb{N}_0)^n}(-1)^{|p|}
  \frac{\delta_{|\!|p|\!|,n}|p|!}{p_1!\cdots p_n!}\,a_{1,\eps}^{p_1}\cdots a_{n,\eps}^{p_n},\qquad n=1,\ldots,N-1.
\label{bne}
\ee
The summation over $j$ yields a factor of the form
\be
 \sum_{j=0}^{N-1}\omega^{j(N-1+i-\ell-\ell'+n)}=\begin{dcases} N,\ & N-1+i-\ell-\ell'+n\equiv 0\ \ (\mathrm{mod}\ N),\\[.15cm]
  0,\ & N-1+i-\ell-\ell'+n\not\equiv 0\ \ (\mathrm{mod}\ N), \end{dcases}
\ee
and since $N-1+i-\ell-\ell'+n>-N$, it follows that the $T_{i,\eps}$-coefficients to $\eps^m$ for $m$ negative are $0$.
The limit $\eps\to0$ is therefore well-defined, resulting in
\be
 T_i=\frac{1}{2k_{N-1}}\sum_{\ell,\ell',n=0}^{N-1}b_{n}\kappa_{ab}(J_{\ell}^aJ_{\ell'}^b)\delta_{N-1+i-\ell-\ell'+n,0},
\ee
whose nonzero terms are seen to match the expression in (\ref{Tifinal}).

For the central parameters, we evaluate
\begin{align}
 c_{i,\eps}&=\eps^i\sum_{j=0}^{N-1}\omega^{ij}c_{(j)}
  =\eps^i\sum_{j=0}^{N-1}\omega^{ij}\frac{\sum_{\ell=0}^{N-1}\omega^{-j\ell}\eps^{-\ell}k_{\ell,\eps}\dim\g}{\sum_{\ell'=0}^{N-1}
   \omega^{-j\ell'}\eps^{-\ell'}k_{\ell',\eps}+Nh^\vee}
 \nonumber\\[.15cm]
 &=\frac{\dim\g}{k_{N-1,\eps}}\sum_{\ell,n=0}^{N-1}b_{n,\eps}k_{\ell,\eps}\sum_{j=0}^{N-1}(\omega^j\eps)^{N-1+i-\ell+n}
  +\mathcal{O}(\eps^{i+1}),
\end{align}
from which it follows that
\be
 c_i=\frac{\dim\g}{k_{N-1}}\sum_{\ell,n=0}^{N-1}b_{n}k_{\ell}\delta_{N-1+i-\ell+n,0}
  =N\dim\g\,\delta_{i,0},
\ee
again confirming (\ref{cSug}).

\section{Galilean $W_3$ algebras}
\label{Sec:GW3}

Higher-order Galilean contractions can also be applied to W-algebras. Below, we present the results for the $W_{3}$ algebra.

\subsection{$W_3$ algebra}
\label{Sec:W3}

The $W_3$ algebra \cite{Zam85} of central charge $c$
is generated by a Virasoro field $T$ and a primary field $W$ of conformal weight $3$, with star relations
\be
 T\ast T\simeq\tfrac{c}{2}\{\I\}+2\{T\},\qquad T\ast W\simeq3\{W\},\qquad 
 W\ast W\simeq\tfrac{c}{3}\{\I\}+2\{T\}+\tfrac{32}{22+5c}\{\Lambda^{2,2}\},
\ee
where
\be
 \Lambda^{2,2}=(TT)-\tfrac{3}{10}\pa^2T ,
\ee
is quasi-primary.

\subsection{Galilean $W_3$ algebra of order $2$}
\label{Sec:W3o2}

Following \cite{ChrisThesis,RR17}, we now recall the structure of the 
second-order Galilean $W_3$ algebra \cite{ABFGR13,GMPT13,CGOR16}.
It is generated by the four fields $T_0,T_1,W_0,W_1$, 
with central parameters $c_0$ and $c_{1}$, and nontrivial star relations
\be
 T_i\ast T_j\simeq\tfrac{c_{i+j}}{2}\{\I\}+2\{T_{i+j}\},\qquad T_i\ast W_j\simeq 3\{W_{i+j}\},\qquad i+j\in\{0,1\},
\ee
and
\begin{align}
W_0\ast W_0\simeq\tfrac{c_{0}}{3}\{\I\}+2\{T_0\}+\tfrac{64}{5c_1}\{\La^{2,2}_{0,1}\}-\tfrac{32(44+5c_0)}{25c_1^2}\{ \La_{1,1}^{2,2}\},\qquad
W_0\ast W_1\simeq \tfrac{c_1}{3}\{\I\} + 2\{T_1\} + \tfrac{32}{5c_1}\{\La^{2,2}_{1,1} \},
\end{align}
where 
\be
\La^{2,2}_{0,1}=(T_0T_1)-\tfrac{3}{10}\pa^{2}T_{1}, \qquad \La^{2,2}_{1,1}=(T_1T_1)
\ee
are quasi-primary. We note that a nonzero $c_1$ can be scaled away by renormalising as
$T_1,W_1\to\hat{T}_1=\frac{T_1}{c_1},\hat{W}_1=\frac{W_1}{c_1}$.

\subsection{Infinite hierarchy}
\label{Sec:W3General}

For any $N\in\mathbb{N}$, the algebra $W_{3}^{\otimes N}$ is generated by the $2N$ fields $\{T_{(i)},W_{(i)}\,|\,i=0,\ldots,N-1\}$, 
and has central charges $\{c_{(i)}\,|\,i=0,\ldots,N-1\}$.
As outlined in the following, the corresponding Galilean algebra is well-defined.
In tune with the general prescription in Section \ref{Sec:Higher}, we thus confirm that
the $N$th-order Galilean $W_3$ algebra $(W_3)_G^N$ is generated by the fields $\{T_i,W_i\,|\,i=0,\ldots,N-1\}$ and has 
central parameters $\{c_i\,|\,i=0,\ldots,N-1\}$.

First, it straightforwardly follows that
\be
 T_i\ast T_j\simeq\tfrac{c_{i+j}}{2}\{\I\}+2\{T_{i+j}\},\qquad T_i\ast W_j\simeq 3\{W_{i+j}\},\qquad i+j\in\{0,\ldots,N-1\},
\label{QPrels}
\ee
while
\be
 T_i\ast T_j\simeq T_i\ast W_j\simeq W_i\ast W_j\simeq0,\qquad i+j\geq N.
\label{TTT}
\ee
To determine $W_{i}\ast W_{j}$ in $(W_3)_G^N$ for $i+j=0,\ldots,N-1$, 
we compute the corresponding star relation $W_{i,\eps}\ast W_{j,\eps}$ in $W_{3}^{\otimes N}$,
\begin{align}
 W_{i,\eps}\ast W_{j,\eps}
 &=\eps^{i+j}\sum_{r,s=0}^{N-1}\omega^{ir+js}\,W_{(r)}\ast W_{(s)}
 \nn
 &\simeq\eps^{i+j}\sum_{r=0}^{N-1}\omega^{(i+j)r}\Big[\frac{c_{(r)}}{3}\{\I\}+2\{T_{(r)}\}+\frac{32}{22+5c_{(r)}}\{\Lambda^{2,2}_{(r)}\}\Big]
 \nn
 &=\frac{c_{i+j,\eps}}{3}\{\I\}+2\{T_{i+j,\eps}\}+\eps^{i+j}\sum_{r=0}^{N-1}\frac{32}{22+5c_{(r)}}\omega^{(i+j)r}\{\Lambda^{2,2}_{(r)}\}.
\end{align}
Recycling the expansion techniques of Section \ref{Sec:GenSug}, we find that
\begin{align}
 \sum_{r=0}^{N-1}\frac{32}{22+5c_{(r)}}(\omega^r\eps)^{i+j}\Lambda^{2,2}_{(r)}
 &=\frac{32}{5Nc_{N-1,\eps}}\sum_{n,\ell,\ell'=0}^{N-1}b_{n,\eps}
  \sum_{r=0}^{N-1}(\omega^r\eps)^{N-1+i+j-\ell-\ell'+n}(T_{\ell,\eps}T_{\ell',\eps})
 \nonumber\\[.15cm]
 &-\frac{48}{25c_{N-1,\eps}}\sum_{n,\ell=0}^{N-1}b_{n,\eps}\sum_{r=0}^{N-1}(\omega^r\eps)^{N-1+i+j-\ell+n}\pa^2T_{\ell,\eps}
 +\mathcal{O}(\eps^{i+j+1}),
\end{align}
where $b_{n,\eps}$ (and $b_{n}$ appearing in (\ref{sum32}) below) are given as in (\ref{bne}) (respectively (\ref{bn})), but now based on
\be
 a_{m,\eps}=\frac{c_{N-1-m,\eps}+\frac{22N}{5}\delta_{m,N-1}}{c_{N-1,\eps}},\qquad 
 a_{m}=\frac{c_{N-1-m}+\frac{22N}{5}\delta_{m,N-1}}{c_{N-1}},\qquad m=1,\ldots,N-1.
\label{amN}
\ee
In the limit $\eps\to0$, this yields
\be
 \sum_{r=0}^{N-1}\frac{32}{22+5c_{(r)}}(\omega^r\eps)^{i+j}\Lambda^{2,2}_{(r)}
 \to\!\sum_{n=0}^{N-1-i-j}\frac{32b_n}{5c_{N-1}}\!\sum_{t=0}^{N-1-i-j-n}(T_{i+j+n+t}T_{N-1-t})
  -\frac{48N}{25c_{N-1}}\pa^2T_{N-1}\delta_{i,0}\delta_{j,0}.
\label{sum32}
\ee
Observing that, for every pair $r,s\in\{0,\ldots,N-1\}$ such that $r+s\in\{N-1,\ldots,2N-2\}$,
\be
 \Lambda_{r,s}^{2,2}=(T_rT_s)-\frac{3}{10}\pa^2T_{N-1}\delta_{r+s,N-1}
\ee
is a quasi-primary field with respect to $T_0$, we then conclude that, for $i+j\in\{0,\ldots,N-1\}$,
\be
 W_i\ast W_j\simeq\frac{c_{i+j}}{3}\{\I\}+2\{T_{i+j}\}
  +\sum_{n=0}^{N-1-i-j}\frac{32b_n}{5c_{N-1}}\!\sum_{t=0}^{N-1-i-j-n}\{\Lambda_{i+j+n+t,N-1-t}^{2,2}\}.
\ee
Using that $\Lambda_{r,s}^{2,2}=\Lambda_{s,r}^{2,2}$, this can be written as
\begin{align}
 W_i\ast W_j&\simeq\frac{c_{i+j}}{3}\{\I\}+2\{T_{i+j}\}
 \nonumber\\[.15cm]
 &+\sum_{n=0}^{N-1-i-j}\frac{32b_n}{5c_{N-1}}\left(
   \sum_{t=0}^{\lfloor\frac{N-2-i-j-n}{2}\rfloor}2\{\Lambda_{i+j+n+t,N-1-t}^{2,2}\}
   +\{\Lambda_{\frac{N-1+i+j+n}{2},\frac{N-1+i+j+n}{2}}^{2,2}\}\right),
\end{align}
where the last term is present only if $\frac{N-1+i+j+n}{2}$ is integer.

Let us illustrate our findings by summarising the nontrivial star relations for the third-order Galilean algebra $(W_3)^3_G$:
The six generating fields $T_0,T_1,T_2,W_0,W_1,W_2$ satisfy (\ref{QPrels})-(\ref{TTT}) with $N=3$
as well as
\begin{align}
 W_0\ast W_0&\simeq
  \tfrac{c_0}{3}\{\I\}
  +2\{T_0\}
  +\tfrac{64}{5c_2}\{\La^{2,2}_{0,2}\}
  +\tfrac{32}{5c_2}\{\La^{2,2}_{1,1}\}
  -\tfrac{64c_1}{5(c_2)^2}\{\La^{2,2}_{1,2}\}
  -\tfrac{32[(66+5c_0)c_2-5(c_1)^2]}{25(c_2)^3}\{\La^{2,2}_{2,2}\},
\\[.15cm]
 W_0\ast W_1&\simeq
  \tfrac{c_1}{3}\{\I\}
  +2\{T_1\}
  +\tfrac{64}{5c_2}\{\La^{2,2}_{1,2}\}
  -\tfrac{32c_1}{5(c_2)^2}\{\La^{2,2}_{2,2}\},
\\[.15cm]
 W_0\ast W_2&\simeq W_1\ast W_1\simeq
  \tfrac{c_2}{3}\{\I\}
  +2\{T_2\}
  +\tfrac{32}{5c_2}\{\La^{2,2}_{2,2}\},
\end{align}
where
\be
 \La^{2,2}_{0,2}=(T_0T_2)-\tfrac{3}{10}\pa^2T_2,\qquad
 \La^{2,2}_{1,1}=(T_1T_1)-\tfrac{3}{10}\pa^2T_2,\qquad
 \La^{2,2}_{1,2}=(T_1T_2),\qquad
 \La^{2,2}_{2,2}=(T_2T_2)
\ee
are quasi-primary.

\subsection{Renormalisation}
\label{Sec:Ren}

We now consider $(W_3)_G^N$ in the special case where
\be
 c_i=c^i,\qquad i=1,\ldots, N-1,
\label{cc}
\ee
for some $c\in\mathbb{C}^\times$, leaving only two independent central parameters: the central charge $c_0$ and $c$.
The $a_m$ coefficients in (\ref{amN}) then simplify to
\be
 a_m=c^{-m}\big(1+\big[c_0+\tfrac{22N}{5}-1\big]\delta_{m,N-1}\big),\qquad m=1,\ldots,N-1.
\ee
Correspondingly, the inverse of the matrix $A$ in (\ref{Aeta}) is given by
\be
 A^{-1}=I-c^{-1}\eta+\big[1-c_0-\tfrac{22N}{5}\big](c^{-1}\eta)^{N-1},
\ee
so (for $N>2$)
\be
 b_0=1,\qquad b_1=-c^{-1},\qquad b_n=0\ \ (1<n<N-1),\qquad
 b_{N-1}=\big[1-c_0-\tfrac{22N}{5}\big]c^{-(N-1)}.
\ee
Let us also introduce the renormalised generators
\be
 \Th_i=c^{-i}T_i,\qquad \Wh_i=c^{-i}W_i,\qquad i=0,\ldots, N-1,
\ee
and ditto quasi-primary fields
\be
 \Lh_{r,s}^{2,2}=c^{-r-s}\La_{r,s}^{2,2}.
\ee
In terms of these, the nontrivial star relations are given by ($i+j\in\{0,\ldots,N-1\}$)
\be
 \Th_i\ast\Th_j\simeq\frac{c_0^{\delta_{i+j,0}}}{2}\{\I\}+2\{\Th_{i+j}\},\qquad 
 \Th_i\ast\Wh_j\simeq 3\{\Wh_{i+j}\},
\ee
and
\begin{align}
 \Wh_i\ast\Wh_j&\simeq\frac{c_0^{\delta_{i+j,0}}}{3}\{\I\}+2\{\Th_{i+j}\}
 +\tfrac{32}{5}\big[1-c_0-\tfrac{22N}{5}\big]\{\Lh_{N-1,N-1}^{2,2}\}\delta_{i+j,0}
 \nonumber\\[.15cm]
 &+\tfrac{32}{5}\sum_{n=0,1}\sum_{t=0}^{N-1-i-j-n}(-1)^n\{\Lh_{i+j+n+t,N-1-t}^{2,2}\}.
\end{align}
The central parameter $c$ has thus been absorbed by a renormalisation of the algebra generators.

A similar absorption is also possible in the Galilean Sugawara construction of Section \ref{Sec:GenSug}, 
with
\be
 \Jh_i^a=k^{-i}J_i^a,\qquad \Th_i=k^{-i}T_i,\qquad i=0,\ldots,N-1,
\ee
where $k_i=k^i$, $i=1,\ldots,N-1$, for some $k\in\mathbb{C}^\times$.
The renormalised Galilean Virasoro generators are then given by
\be
 \Th_i=\tfrac{1}{2}\sum_{n=0,1}\sum_{t=0}^{N-1-i-n}\kappa_{ab}(\Jh_{i+n+t}^a\Jh_{N-1-t}^b)
  +\tfrac{1}{2}[1-k_0-Nh^\vee]\kappa_{ab}(\Jh_{N-1}^a\Jh_{N-1}^b)\delta_{i,0},
\ee
while the nontrivial star relations read ($i+j\in\{0,\ldots,N-1\}$)
\be
 \Jh^a_i\ast\Jh^b_j\simeq\kappa^{ab}k_0^{\delta_{i+j,0}}\{\I\}+{f^{ab}}_c\{\Jh^c_{i+j}\},
 \quad
 \Th_i\ast\Jh_j^a\simeq\{\Jh_{i+j}^a\},
 \quad
 \Th_i\ast\Th_j\simeq\tfrac{N\dim\g}{2}\{\I\}\delta_{i+j,0}+2\{\Th_{i+j}\}.
\ee

\section{Discussion}
\label{Sec:Discussion}

In our continued exploration \cite{ChrisThesis,RR17} of Galilean contractions, we have presented a generalisation of the 
contraction prescription to allow for inputs of any finite number of OPAs or vertex algebras. This has resulted in
hierarchies of higher-order Galilean conformal algebras, including Virasoro, affine Kac-Moody and $W_{3}$ algebras.

Asymmetric Galilean $N=1$ superconformal algebras, corresponding to an $N=(1,0)$ supersymmetry, 
can be obtained \cite{BDMT14,BDMT17,CGOR16,BJLMN16} from a Galilean contraction of
the tensor product, $S\mathfrak{Vir}\otimes\Vir$, of an $N=1$ superconformal algebra, $S\mathfrak{Vir}$, and the Virasoro algebra. 
As we hope to discuss in detail elsewhere, this extends to contractions of a conformal symmetry algebra with any subalgebra thereof. 
For example, one readily generalises our contraction prescription to the asymmetric tensor product $W_3\otimes\Vir$, 
where one contracts the Virasoro subalgebra of $W_{3}$ with a separate Virasoro algebra.
This yields an OPA generated by fields $T_{0}, T_{1}, W$, with nonzero star relations ($i + j \in \{ 0,1 \}$)
\be
T_{i} \ast T_{j}\simeq \tfrac{c_{i+j}}{2} \{ \I \} + 2 \{ T_{i+j} \},\qquad
T_{0} \ast W \simeq 3 \{ W \},\qquad
W \ast W \simeq \tfrac{c_{1}}{3} \{ \I \} + 2 \{ T_{1} \}+ \tfrac{32}{5c_{1}} \{ \Lambda_{1,1}^{2,2} \} .
\ee
There is significant freedom in such contractions, leading to a variety of inequivalent Galilean algebras.

Other avenues for future research include representation theory and free-field realisations.
The representation theory of the Galilean Virasoro algebra, also known as the $W(2,2)$ algebra,
has already been studied in some detail \cite{ZD07,LGZ08,LS08,WL09,BBGR14,AR16}.
In general, though, the representation theory of Galilean algebras remains largely undeveloped and is
entirely unexplored in the case of the higher-order algebras introduced in the present note. 

Free-field realisations \cite{DF84,Wak86,FMS86,FL88,FF90,BS93,Fre94,Ras96,Ras98,Kau00}
have been central to many developments in and applications of conformal field theory, and it seems 
natural to expect that free fields will play a similar role when Galilean conformal symmetries are present.
This includes the representation theory of the Galilean algebras alluded to above.
Although realisations of the Galilean Virasoro algebra and some of its superconformal extensions have been considered 
\cite{BJMN16,AR16,BJLMN16}, a systematic approach and general results are still lacking.

\subsection*{Acknowledgements}

JR was supported by the Australian Research Council under the Discovery Project scheme, project number DP160101376. 
CR was funded by a University of Queensland Research Scholarship.
The authors thank David Ridout for helpful discussions.

%

\end{document}